\documentstyle[12pt,epsf]{article}

\begin{document}

\begin{center} 

{\Large \bf What can we learn by probing Trans-Planckian physics}\footnote{Talk given at CICHEP-01, Cairo, Jan. 2001} \\
\end{center}

\begin{center}
M.  Bastero-Gil  \\

{\small \it Scuola Normale Superiore \\ Pzza. dei Cavalieri n. 7, 56126-Pisa (Italy)  }
\end{center}








{ \small In this talk we address the issue of how the
observables in our present Universe are affected by processes that may
have occured at superplanckian energies (referred to as the
{\it transplanckian regime}). For example, the origin of the
cosmological perturbation spectrum. We model the transplanckian regime by
introducing a 1-parameter family of smooth non-linear 
dispersion relations which modify the frequencies at very short
distances.  
For this family of dispersions, we present the exact solutions and
show that the CMBR spectrum is that 
of a (nearly) black body, and that the adiabatic vacuum is the only
choice for the initial conditions. 
A particular feature of the family of dispersion functions
chosen  is the production of 
ultralow frequencies at very high momenta $k$ (for $k>M_P$). Modes
with ultralow frequencies equal or less than the current 
Hubble rate are still frozen today. Therefore, their energy today provides a
strong candidate for the dark energy of the Universe. 
}

\section{Introduction}

Nowadays, quantum field theory is view as an effective description of
collective degrees of freedom valid below some cutoff scale. If not
other, in particle physics the natural cutoff scale is provided by the
Planck mass, near which quantum gravity effects will become important.
In the definition of an effective field theory, it is implicit that the
long wavelength phenomena are decoupled from the small scale
processes, such that we do not need to know about the short distance
behavior of the underlying fundamental theory in order to study the
long distance 
description. However, there are cases where 
this long/short distance physics separation breaks down. For example,
when studying the origin of the Hawking radiation in Black Hole
physics; or when studying the spectrum of primordial cosmological
perturbations generated during inflation. In both cases the physical
momentum gets blue-shifted back in time, such that the low energy
modes evolve from degrees of freedom above the cutoff. Therefore, 
the effective field theory approach is not viable due to 
the presence of a strong redshift which mixes the ultraviolet and infrared
regimes.

The problem was first raised in Black Hole physics, trying to
explain the origin of Hawking radiation. 
In a series of papers~\cite{jacobson,unruh,brout}, it
was demonstrated that the Hawking radiation remains 
unaffected by modifications of the ultra high energy regime, expressed
through the modification of the usual linear dispersion relation at
energies larger than a certain ultraviolet scale $k_C$. In particular,
following the sonic black hole analogy, Unruh~\cite{unruh} proposed a dipersion
relation that goes asymtotically constant,
\begin{equation}
\omega(k) = k_C \tanh^{1/n} \left[(k/k_C)^n \right] \,,
\label{dunruh}
\end{equation}
whilst Corley and Jacobson~\cite{jacobson} adopted the function
\begin{equation}
\omega(k)= k^2 ( 1 - k^2/k_C^2 ) \,,
\label{dcorley}
\end{equation}
which they considered as the lowest order term in the
derivative expansion of a generic dispersion relation.

If we think in terms of waves propagating in an inhomogeneous medium, 
it is reasonable to assume that the dispersion relation for the mode
propagation will get modified when the  mode start probing the
underlying structure of the background; in our case, the
transplanckian regime. But we lack a fundamental theory, valid at all
energies, able to describe the transition. At the same time, this
makes the model building of the transplanckian regime very
interesting.  The main issue is how much are the known
observables affected by the unknown theory. 
The apparently {\it ad hoc} modification of the dispersion relation
introduced at high energies is constrained by the criterion that its low energy
predictions do no conflict the observables. In the case of an
expanding Friedmann-Lemaitre-Robertson-Walker (FLRW) spacetime, one 
can ask wether the standard predictions of inflation are or not sensitive
to trans-planckian physics. Martin and Brandenberger in
Ref.~\cite{brand} (see also~\cite{niemeyer,kowalski}) studied
this adopting the above dispersion relations, Eqs. (\ref{dunruh}) and
 (\ref{dcorley}), and found that indeed different dispersion
relations lead to different results for the CMBR spectrum. Deviation
from the standard scale invariant spectrum were obtained 
when using Corley and Jacobson's 
dispersion function, which is not well defined for $k \gg k_C$. They
conjecture that the observed power spectrum can always be recovered
 by using a smooth dispersion relation, which ensures an adiabatic
time-evolution of the modes.    

In Refs.~\cite{brand,niemeyer} the authors have also 
demonstrated that the  problem of calculating the spectrum of
perturbations with a time-dependent dispersive frequency 
can be reduced to the familiar topic of particle creation on a
time-dependent background~\cite{partcreation}.  
In this talk we adopt their method in studying the trans-planckian
problem, but we introduce a new family of dispersion relations to
model that regime~\cite{laura}. This class of functions has the
following features:  
it is smooth, nearly linear for energies less than the
Planck scale, reaches a maximum, and attenuates to zero at ultrahigh
momenta thereby producing ultralow frequencies at very short
distances (see fig. (\ref{fig1})). This choice of functions is
motivated by superstring duality~\cite{superstrings} (which applies at
transplanckian 
energies). Our family of dispersion relations exhibits $dual$
behavior, i.e., appearance of ultra-low mode frequencies both at low
and high momenta.

Below we present the exact solutions to the mode equation,  
 and the resulting CMBR spectrum. The major
contribution to the CMBR spectrum comes from the long wavelength modes
when they re-enter the horizon. The spectrum is nearly insensitive to
the very short wavelength modes inside the Hubble horizon. Therefore,
by taking the 
frequency dispersion relations to be the general class of Epstein
functions~\cite{epstein}, we check and lend strong support to the 
conjecture made in Ref.~\cite{brand}. 

On the other hand, the family of dipersion chosen present the distinctive
feature of having a ``tail'' of modes with ultralow frequencies, less
or equal to the current Hubble constant $H_0$. It follows that 
the $tail$ modes are still currently frozen. They provide a
unique candidate for the dark energy of the universe (see
Ref.~\cite{laura}).

\section{The Model and CMBR spectrum} 

The initial power spectrum of the metric perturbations can be computed once
we solve the 
time-dependent equations for the scalar and tensor sector. The mode
equations for both sectors reduce~\cite{equations} to a
Klein-Gordon equation of the form
\begin{equation}  
\mu_n^{\prime \prime} + \left[ n^2 - \frac{a^{\prime \prime}}{a} \right]
\mu_n=0 \,,
\label{kg}
\end{equation}
where the prime denotes derivative with respect to conformal time.
Therefore, studying perturbations in a FLRW background is equivalent
to solving the mode equations for a scalar field $\mu$ related
(through Bardeen variables~\cite{equations}) to the
perturbation field  in the expanding
background. The dynamics of the scale factor is determined by the
evolution of the background inflaton field $\phi$, and the Friedmann equation. 
We will consider the class of inflationary scenarios that has a power
law solution for the scale factor $a(\eta)$ in conformal time, 
$a(\eta)=|\eta_c/\eta|^\beta$, with $\beta \ge 1$. 

Eq. (\ref{kg}) represents a linear dispersion relation for the
frequency $\omega$, 
$\omega^2= k^2 = n^2/a^2$. 
This dispersion relation holds for values of
momentum smaller than the Planck scale. There is no reason to believe
that it remains linear at ultra-high energies larger than $M_P$.  
However, we should stress that any modeling of
Planck scale physics even by analogy with already familiar systems is
pure speculation. We lack the fundamental theory that may naturally
motivate or reproduce such dispersive behavior.  

\begin{figure}[t]
\epsfxsize=10cm
\epsfxsize=10cm
\hfil \epsfbox{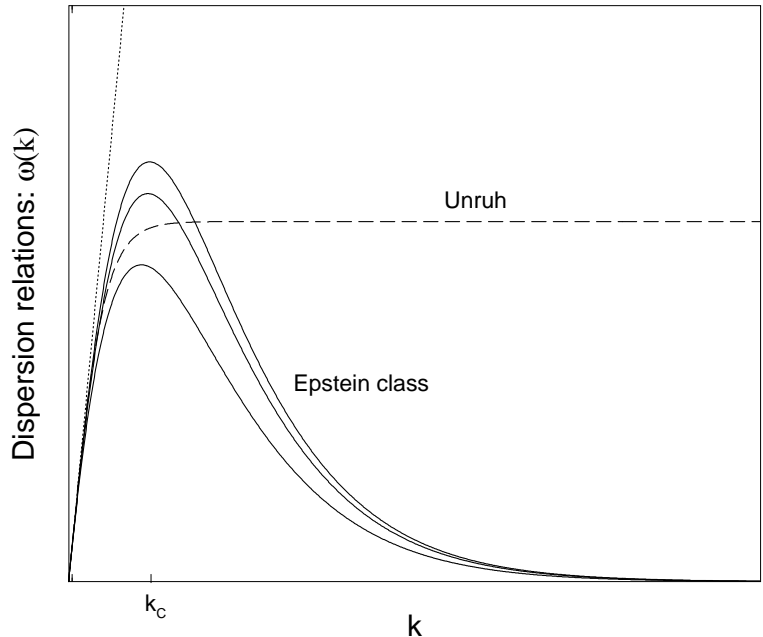} \hfil
\caption{{ Shown is our family of dispersion relations,
for $\beta=1$ and representatives values of $\epsilon_1$ (solid
lines). We have also included the Unruh's 
dispersion relation (dashed line) and the linear one (dotted line) for
comparison.} }
\label{fig1}
\end{figure}

In what follows, we will replace the linear relation with a
nonlinear dispersion relation 
$\omega(k)=F(k)$. Therefore, in Eq. (\ref{kg}), $n^2$ should be
replaced by 
$n_{eff}^2 = a(\eta)^2 F(k)^2 = a(\eta)^2 F[n/a(\eta)]^2$. For future
reference, we also define the generalised comoving frequency as 
$\Omega_n^2 = n_{eff}^2 - a^{\prime \prime}/a$. 
We consider the following
Epstein function~\cite{epstein} for the dispersion relation:
\begin{eqnarray}
\omega^2(k) &=& F^2(k)= k^2 \left(\frac{\epsilon_1}{1+ e^x} +
\frac{\epsilon_2 e^x}{1+e^x} + \frac{\epsilon_3 e^x}{(1+e^x)^2}\right)
\,,\label{omega} \\
n^2_{eff} &=& a^2(\eta) F^2(n,\eta)= n^2 \left(\frac{\epsilon_1}{1+ e^x} +
\frac{\epsilon_2 e^x}{1+e^x} + \frac{\epsilon_3 e^x}{(1+e^x)^2}\right) \,,
\label{disp}
\end{eqnarray}
where $x=(k/k_C)^{1/\beta}=  A |\eta|$, with $
A=(1/|\eta_c|)(n/k_C)^{1/\beta}$. This is the most general expression
for this family of functions.  For our purposes, we will constrain
some of the parameters of the Epstein family in order to satisfy
the features required for the dispersion relation  as
follows.  
First, in order to have
ultralow frequencies for very high momenta, we
demand that the dispersion function goes asymptotically to zero. 
That fixes $\epsilon_2=0$. And the condition of a nearly linear
dispersion relation for 
$k<k_C$ requires that  $ 2 \epsilon_1 + \epsilon_3 = 4$. 
Still we will have a whole family of functions parametrised by the
constant $\epsilon_1$, as can be seen in Fig. \ref{fig1}.

Eq. (\ref{kg}) with the new dispersion function Eq. (\ref{disp}) is exactly
solvable in terms of hypergeometric 
functions~\cite{epstein}. This is a well studied case in the context
of particle creation in a curved background~\cite{partcreation}.  The
contribution from $a^{\prime \prime}/a$ 
is going to be negligible at early times ($\eta \rightarrow -\infty$); 
at late times, it can be absorbed in the dispersion relation
Eq. (\ref{disp}) redefining the constants $\epsilon_i$.  

The correct initial condition is the
vacuum state solution that minimizes the energy~\cite{initial}. The
choice of the 
correct vacuum state is important, since most of the contribution to
the spectrum of perturbations comes from long-wavelength modes. They
are produced at early stages of inflation, being very sensitive to the
initial conditions. 
When $\epsilon_2 \neq 0$, the vacuum state behaves as a
plane wave in the asymptotic limit $\eta \rightarrow -\infty$, with
$\Omega^{(in)}_n \rightarrow \sqrt{\epsilon_2} n$. 
However, when $\epsilon_2=0$ as in our case, 
the correct behavior of the mode function in the
remote past is given by the solution of its evolution equation in the
limit $\eta \rightarrow -\infty$. The exact solution which matches
this asymptotic behavior is then given by:
\begin{equation}
\mu^{(in)}(\eta)= C^{in} \left(\frac{1
+u}{u}\right)^{d}~ _2F_1 [\frac{1}{2}+ d + b,\frac{1}{2}+ d-b, 1+2 d,
\frac{1 +u}{u}] \,, 
\end{equation} 
where $u=exp(A |\eta|)$, $C^{in}$ is a normalization constant, and   
\begin{equation}
b = i \tilde{b} = i \sqrt{ \hat{\epsilon}_1}  \,,\;\;\;\;\;
d = i \tilde{d} = \sqrt{\frac{1}{4} + \hat\epsilon_3} \,,
\end{equation}
where $\hat{\epsilon}_i= (k_C |\eta_c|)^2 (n/k_C)^{2 (1-1/\beta)}
\epsilon_i$. 
At late times, the
solution becomes a squeezed state by mixing of positive and negative
frequencies:
\begin{equation}
\mu_n \rightarrow_{\eta \rightarrow +\infty} \frac{\alpha_n}{\sqrt{2
\Omega^{out}_n}} e^{-i\Omega^{out}_n \eta} + \frac{\beta_n}{\sqrt{2
\Omega^{out}_n}} e^{+i \Omega^{out}_n \eta} \,,
\end{equation}
where $\alpha_n$ and $\beta_n$ are the Bogoliubov coefficients, 
and $\Omega^{out}_n \simeq \sqrt{\epsilon_1} n$; 
$|\beta_n|^2$ gives the
particle creation number per mode $n$. Using the linear transformation
properties of hypergeometric functions~\cite{abramowitz}, 
we find that
\begin{equation}
\left| \frac{\beta_n}{\alpha_n} \right| = e^{-2 \pi 
\tilde{b}} \left|\frac{ \cosh \pi(\tilde{d} + \tilde{b})}{ \cosh
\pi(\tilde{d} - \tilde{b})} \right| \label{betak}\,.
\end{equation}
It is clear from Eq . (\ref{betak})  that the 
spectrum of created particles is nearly thermal to high 
accuracy\footnote{ We mention that we have neglected the  
backreaction effects during the calculation. However, this is
consistent with the result obtained of a
small particle number per mode, in the high momentum regime ($k \gg M_P$) 
and a very small energy contained in these
modes. Because of these results, we do not have the problems mentioned
in Ref.~\cite{tanaka} when discussing trans-planckian physics. For the
energy see Ref.~\cite{laura}.}:   
$
|\beta_n|^2 \simeq e^{-4 \pi \tilde{b}} \,.
$
Thus, we can immediately conclude that the CMBR spectrum is that of a
(nearly) black body spectrum. This means that the spectrum is (nearly)
scale invariant, i.e., the spectral index is $n_s \simeq 1$.
This is consistent with previous results obtained in the
literature~\cite{brand,niemeyer,kowalski}, when using
a smooth dispersion relation and the correct choice of the initial
vacuum state, as discussed above. In Refs.~\cite{brand} and
\cite{niemeyer}, dispersion relations that were originally
applied to black hole physics~\cite{jacobson,unruh} were
used in the context of cosmology. New models of dispersion relations
were proposed by the authors of Refs.~\cite{kowalski}.
Our proposal for a 1-parameter class of models has a significantly
different feature from the above, namely:
the appearence of ultra-low frequency
modes in the transplanckian regime. The implications of such a behavior
for high momenta on the production of dark energy are discussed in
Ref.~\cite{laura}. 
 \section*{Acknowledgments}
We thank A. Riotto, R. Barbieri, R. Rattazzi, L. Pilo,
S. Carroll, C. T. Hill for helpful discussions.

\end{document}